\documentclass[pre, twocolumn, superscriptaddress,showpacs]{revtex4-1}

\usepackage{graphicx}
\usepackage{float}
\usepackage{amsmath}
\usepackage{amssymb}
\usepackage{latexsym}
\usepackage{comment}
\usepackage{bm}
\usepackage{gensymb}
\usepackage{hyperref}
\usepackage{xcolor}
\hypersetup{colorlinks,urlcolor=blue}
\usepackage[most]{tcolorbox}
\usepackage[font=small,labelfont=bf,
   justification=centerlast ]{caption} 
\usepackage{subcaption}
\usepackage{csquotes}
\usepackage{amssymb}
\usepackage{tikz}
\usetikzlibrary{shapes,snakes}

\graphicspath{{Figures/}}

\begin{document}
\title{Columnar structures of soft spheres: Metastability and hysteresis}
\date{\today}
\author{A. Mughal}
\affiliation{Institute of Mathematics, Physics and Computer Science, Aberystwyth University, Penglais, Aberystwyth, Ceredigion, Wales, SY23}
\author{J. Winkelmann} 
\affiliation{School of Physics, Trinity College Dublin, University of Dublin, Dublin 2, Ireland}
\author{D. Weaire} 
\affiliation{School of Physics, Trinity College Dublin, University of Dublin, Dublin 2, Ireland}
\author{S. Hutzler} 
\affiliation{School of Physics, Trinity College Dublin, University of Dublin, Dublin 2, Ireland}

\begin{abstract}
Previously we reported on the stable (i.e. minimal enthalpy) structures of soft monodisperse spheres in a long cylindrical channel. Here, we present further simulations, which significantly extend the original phase diagram up to $D/d=2.714$ (ratio of cylinder and sphere diameters), where the nature of densest sphere packing changes.
However, macroscopic systems of this kind are not confined to the ideal
equilibrium states of this diagram. Consequently, we explore some of the
structural transitions to be expected as experimental conditions are
varied; these are in general hysteretic. We represent these transitions in
a stability diagram for a representative case. Illustrative videos are
included in the Supplemental Material.
\end{abstract}

\maketitle

\section{introduction}

\begin{figure}[h!]
\begin{center}
\includegraphics[width=0.75\columnwidth]{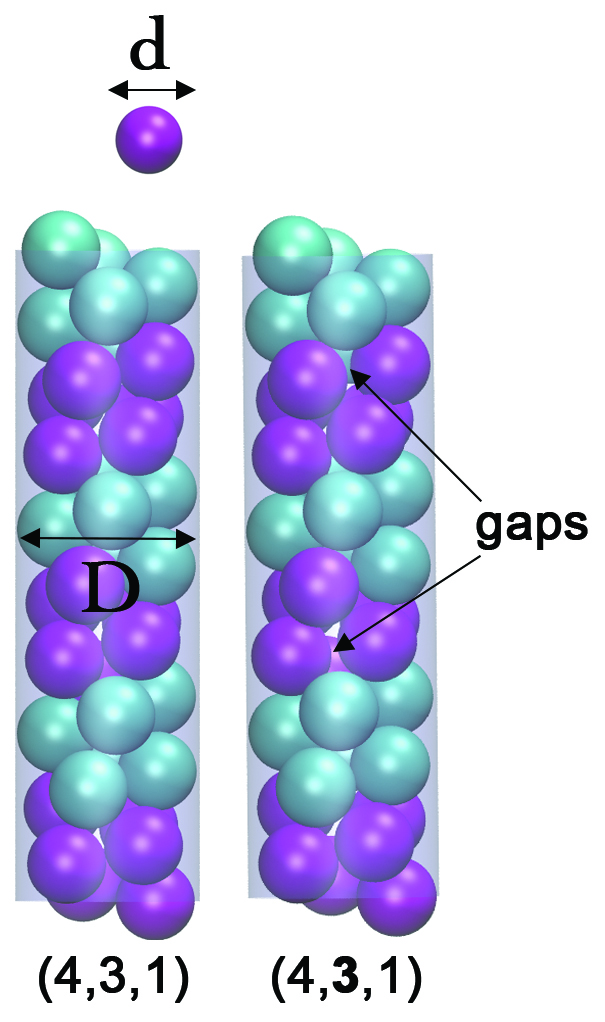}
\caption{Two examples of  columnar crystals,  generated by packing hard
spheres of diameter $d$ into a cylinder of diameter $D$. The structure
shown on the left is the \emph{uniform} packing $(4,3,1)$ in phyllotactic notation
(see Ref.{}\cite{mughal2012dense}), with every sphere having a coordination
number $z=6$. The structure on the right is the related \emph{line-slip} structure
$(4,{\bf 3},1)$. The gaps indicated in the second structure correspond to a
loss of contacts when compared to the first structure.
}
\label{spherepacking}
\end{center}
\end{figure}

There is an increasing interest in the field of columnar structures, two examples of which are shown in Fig.{}\ref{spherepacking}.
These range from packings of bubbles \cite{winkelmann2017simulation, meagher2015experimental,
saadatfar2008ordered, pittet1996structural, Tobin2011,
boltenhagen1998giant, boltenhagen1998structural} and colloidal particles \cite{lohr2010helical, fu2017assembly, yin2003self,
wu2017confined} in cylinders at the macroscopic level, to molecules or
particles in nanotubes \cite{barzegar2015csub60, khlobystov2005molecules, Troche2005},
polymer coated nanoparticles in cylinders \cite{sanwaria2014helical}, as
well as similar packings {\em on} cylinders \cite{Harvard2018}. 
Another stimulating discovery is that of tubular crystals \cite{douglass2017stabilization}, which have been found in simulations:
these are made up of interlocking columnar structures \cite{mughal2011phyllotactic, mughal2012dense, chan2011densest, fu2017assembly, mughal2013screw, fu2016hard, mughal2014theory, wood2013self}.
All of the above work is informed by extensive computational studies of cylindrical dense packings and optimal packings for soft spheres and atoms interacting with pair potentials.

A recent development offers a new experimental method for the fabrication of cylindrical packings of polymer beads by the use of a lathe to induce a centripetal force \cite{Koreans2017}.

We have recently presented an analysis of equilibrium structures of soft
spheres as a function of pressure $P$ and the diameter ratio $D/d$ of a
cylinder ($D$) and the soft spheres of equal size ($d$) that are packed
within it \cite{winkelmann2017simulation}. These simulation results were
for structures which were found to have the lowest enthalpy. It was
recognised that these might be of limited value for macroscopic
experiments. Accordingly we have now undertaken an exploration of
metastability and hysteresis. That is, we ask the question: Given a certain
structure, locally stable, and some (experimental) protocol for the
continuous variation of $P$ and/or $D/d$, what sequence of transitions is to be expected?

It is practically impossible to develop a fully comprehensive answer to
this question, given the richness of possibilities presented in Sec.{} 
\ref{sec:phasediagram}.
Indeed, this does not exhaust the
possible list of structures which might conceivably arise in the present
investigation. Here we give a description of the methodology used to
investigate transitions between columnar structures.

Before presenting these new results in Sec.{}\ref{metastability} we will introduce the model (Sec.{}\ref{model}) and present a now greatly expanded phase diagram in Sec.{}\ref{sec:phasediagram}. 

\section{Model and Method}
\label{model}

As described previously \cite{winkelmann2017simulation}, the model that we
adopt consists of spheres of diameter $d$, whose overlap $\delta_{ij}$
leads to an increase in energy $E^S_{ij}$ according to
\begin{equation}
E^S_{ij} = \frac{1}{2} k \delta_{ij}^2\,.
\end{equation}
Here the overlap between two spheres $i$ and $j$ is defined as $\delta_{ij} = \vert \vec{r}_i - \vec{r}_j \vert - d$, where $\vec{r}_i=(r_i, \theta_i, z_i)$ and $\vec{r}_j=(r_j,\theta_j, z_j)$ are the centres of two contacting spheres in cylindrical polar coordinates. A harmonic energy term $E^B_i= \frac{1}{2} k ((D/2 - r_i)-d/2)^2$ accounts for the overlap between the $i$th sphere and the cylindrical boundary.
The spring constant $k$ determines the softness of the spheres.

This is a generic model offering a qualitative and semi-quantitative
understanding of a variety of physical systems.

We conduct simulations using a unit cell of length $L$ (volume
$V=\pi(D/2)^2L$), containing $N$ spheres.
On both ends of the unit cell we impose \emph{twisted} periodic boundary
conditions to represent  a periodic columnar structure of soft spheres.
The periodic boundaries are implemented by placing image spheres above and below the unit cell, where each sphere of the unit cell is moved in $z$-direction by $L$ (and $-L$) and twisted by an angle $\alpha$ (and $-\alpha$, respectively) in the $xy$-plane (for more details see Ref.{}\cite{mughal2012dense}).

Stable structures are found by minimising the enthalpy $H = E + PV$ for a system of $N$ soft spheres in the unit cell, where $E = E^S + E^B$ is the internal energy due to overlaps as described before (see Ref.{}\cite{winkelmann2017simulation}) and $P$ is the pressure.
During the minimisation, the free parameters are the sphere centres
$\{\vec{r}_i\}$, the twist angle $\alpha$, and the length of the unit cell
$L$, while the pressure $P$ is kept constant.
Thus, all simulations are performed at constant pressure and variable volume.

Two different types of minimisation routine are used to minimise the enthalpy $H(\{r_i\}, \alpha, L)$.
The stochastic basin-hopping algorithm \cite{basinhopping} performs a
general search for the {\it global} minimum and the BFGS method \cite{BFGS}
is used for a conjugate gradient algorithm to search for the nearest {\it
local} minimum of enthalpy.

Enthalpy and pressure have to be rescaled accordingly to obtain non-dimensional quantities.
We use the dimensionless enthalpy $h = H / (kd^2)$ and dimensionless
pressure $p = P / (k / d)$, where $k$ is the spring constant and $d$ is the
sphere diameter.

In the limit $p \to 0$ we make contact with previous results for densest
packings of hard spheres \cite{mughal2012dense}.

\section{The Phase Diagram}
\label{sec:phasediagram}

\begin{figure*}
\begin{center}
\includegraphics[width=0.895\textwidth]{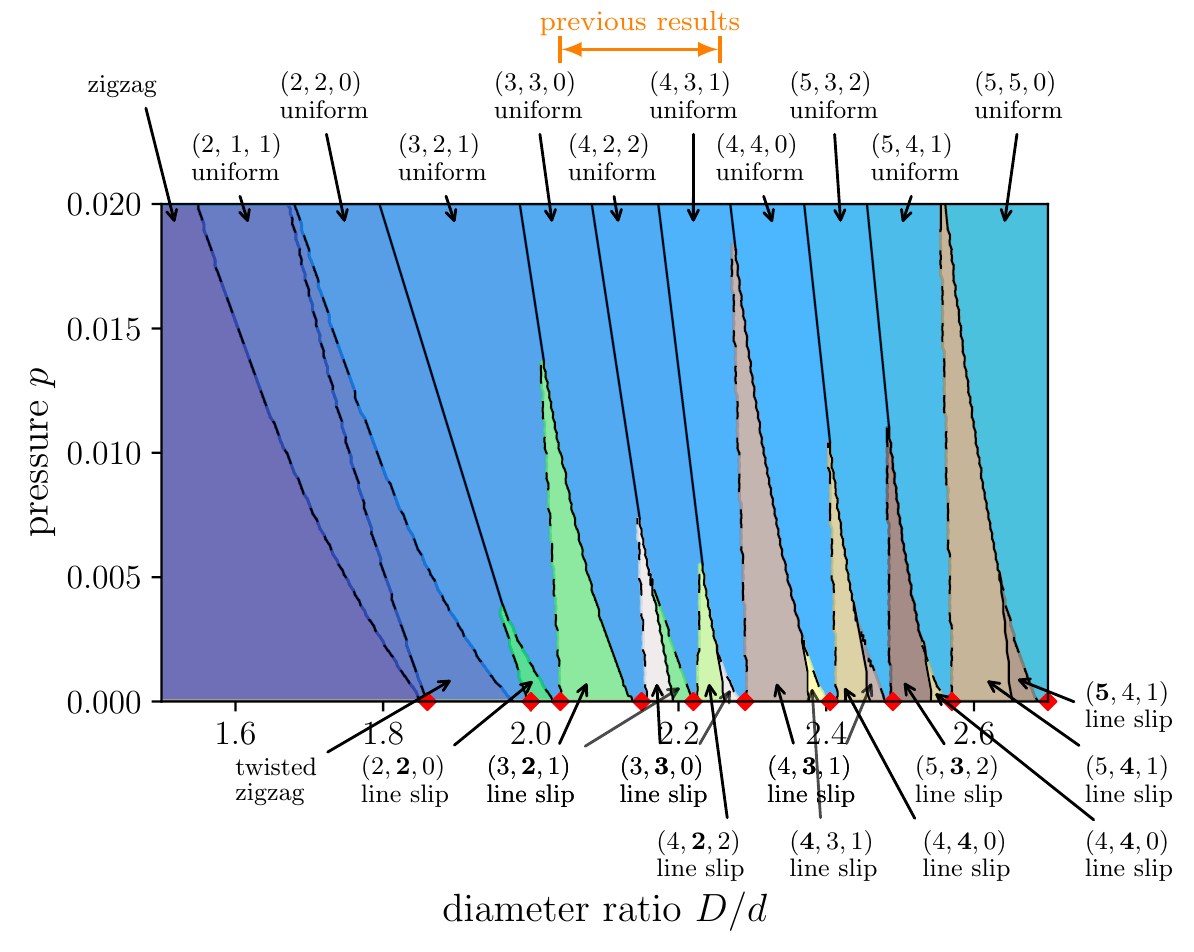}
\caption{Phase diagram for soft sphere packings in cylinders in the range
$1.5\leq D/d\lesssim 2.7$ and dimensionless pressures $p \leq 0.02$. 
The resolution in the diameter ratio of tube to sphere diameter is $\Delta D/ d = 0.0025$ and in the pressure $\Delta p = 0.0004$. Besides the zigzag, the $(2, 1, 1)$ uniform structure, and the twisted zigzag, there are $10$ uniform (blue shaded) and $14$ related line-slip structures (green and brown).
Small regions that contain the $(\bm{2}, 1, 1)$ and the $(\bm{3}, 2, 1)$ line slips, which were found for hard spheres, are not visible in this phase diagram due to the finite resolution. Discontinuous transitions are indicated by solid black lines, while continuous transitions are indicated by black dashed lines.
The diamond symbols at $p = 0$ correspond to the hard sphere uniform close-packed structures \cite{mughal2012dense}.
The range of the previous results published in Ref.{}\cite{winkelmann2017simulation} is indicated by the orange arrow above the diagram.
}
\label{phasediagram}
\end{center}
\end{figure*}

All the structures that we found can be classified either as being {\it
uniform} structures (previously called {\it symmetric} in hard sphere
packings), or {\it line-slip} structures, see Fig.~\ref{spherepacking}.


In Fig.{}\ref{phasediagram} we have greatly extended the range of the
previously published phase diagram to $1.5 \le D/d \lesssim 2.7$ (Fig.{}2 of Ref.{}\cite{winkelmann2017simulation} was only
$2.00 \le D/d \le 2.22$).
The upper limit marks the point at which the character of the optimal structures changes radically; beyond this point the structures contain inner spheres which are not in contact with the cylindrical wall \cite{mughal2012dense}.

The pressure range is limited to $p \leq 0.02$, beyond which line-slip
structures are absent. At  large pressures the model may be regarded as
unrealistic. For a system of bubbles, which was the initial context for
this work, one encounters the \enquote{dry limit}, as $p$ increases.
At this point all liquid in a foam has been removed. 

The same procedure as before was used to obtain the phase diagram:
The simulations were run with $N = 3, 4, 5$ spheres in the unit cell, and for a given value of pressure and diameter ratio the structure with the lowest enthalpy was selected for the phase diagram.

Above the diameter ratio of $2.0$, we find 24 distinct structures. There are 10 uniform packings and the remaining 14 structures are their corresponding line slips. Below $D /  d = 2.0$ we observe the bamboo structure, the zigzag packing, the $(2,1,1)$ uniform arrangement, and the twisted zigzag structure [i.e., the $(2,{\bf 1},1)$ line slip]. 

The transitions between different structures can be classified as follows.
In {\it continuous} phase transitions (marked with dashed lines in Fig. \ref{phasediagram}) a structure transforms smoothly into another by gaining or losing a contact.
This can be observed in supplemental video S0 \cite{supplemental}, which shows an overview over all structures at a low pressure, together with the corresponding rolled-out contact network and the structure's position in the phase diagram.
Both the enthalpy and its first derivative are continuous functions of $D/d$. 
This type of transition is found between a uniform structure and a line
slip. The zigzag to $(2, 1, 1)$ transition is also continuous.
{\it Discontinuous} transitions (solid lines in Fig. \ref{phasediagram})
are transitions where one structure changes abruptly into another, as can also
be seen in video S0 \cite{supplemental}.

\section{Metastability and hysteresis}
\label{metastability}

Whereas the computation of a phase diagram entailed a search for the {\it
global} minimum, involving an algorithm that allowed radical changes of
structure to be explored, here we pursue a more limited objective. {\it
Given a stable structure, possibly metastable, how does it change when the
pressure $p$ and diameter ratio $D/d$ are continuously varied, if it is to
remain in the local minimum of enthalpy?}

We do this by computing trajectories in the $(p, D/d)$ plane,  and
recording boundaries where the structure changes to one of a different
character. With a sufficient number of trajectories, a \textit{stability diagram} is built up, i.e. a map of the location of structural transitions.

With the procedure that we adopt (the conjugate gradient method), structures of high symmetry, such as the $(4, 2, 2)$, can get stuck on an saddle point, where the enthalpy is not minimal, but its gradient is zero.
This can be avoided by applying a {\it small} random perturbation to the
structure, which displaces it from the saddle point, followed by a local minimisation.

The simulation procedure was carried out for a unit cell with $N = 12$
spheres. This choice is commensurable with structures that contain $N =1,
2, 3$ or $4$ spheres in the unit cell, and as such includes all of the
known structures from the previously identified hard-sphere packings (see
Table 1 in Ref.{}\cite{mughal2012dense} for structures and the number of spheres
in the unit cell). An exception to this are packings which contain $N=5$
spheres; however in the hard sphere case these occur only for high values of $D/d$ and as such are out of the range of the present simulations. 

In the following example we consider transitions between the $(3,2,1)$ and $(4,2,2)$ uniform structures and the associated $(3, {\bf 2}, 1)$ line slip. We begin by plotting the changes in enthalpy that occur for a structure when $p$ is held fixed and $D/d$ is allowed to vary steadily. We show that at a low pressures it is possible to start with any one of these structures and continuously transform one structure into another: that is, a change in $D/d$ can transform a uniform structure into a line-slip arrangement by the loss or formation of a contact.
This is accompanied by a smooth variation in the enthalpy, or at most a
change in slope for the derivative of the enthalpy when a new contact is
formed. However, at higher pressures these transitions are no longer
reversible and show evidence of hysteresis, such discontinuous
transformations are accompanied by a discontinuity in enthalpy. From such
results we eventually obtain a stability diagram (as described below in
Sec.{}\ref{ss:stability})
operational for any trajectory taken in terms of $D/d$ and $p$.

In order to appreciate the significance of the enthalpy curves, the reader may wish to refer forward to the relevant stability diagram, Fig.{}\ref{hysteresis321} as well as the phase diagram Fig.{}\ref{phasediagram} and the Supplemental Material \cite{supplemental}.

\subsection{Enthalpy curves}

\begin{figure}[h!]
\begin{center}
\includegraphics[width=0.99\columnwidth]{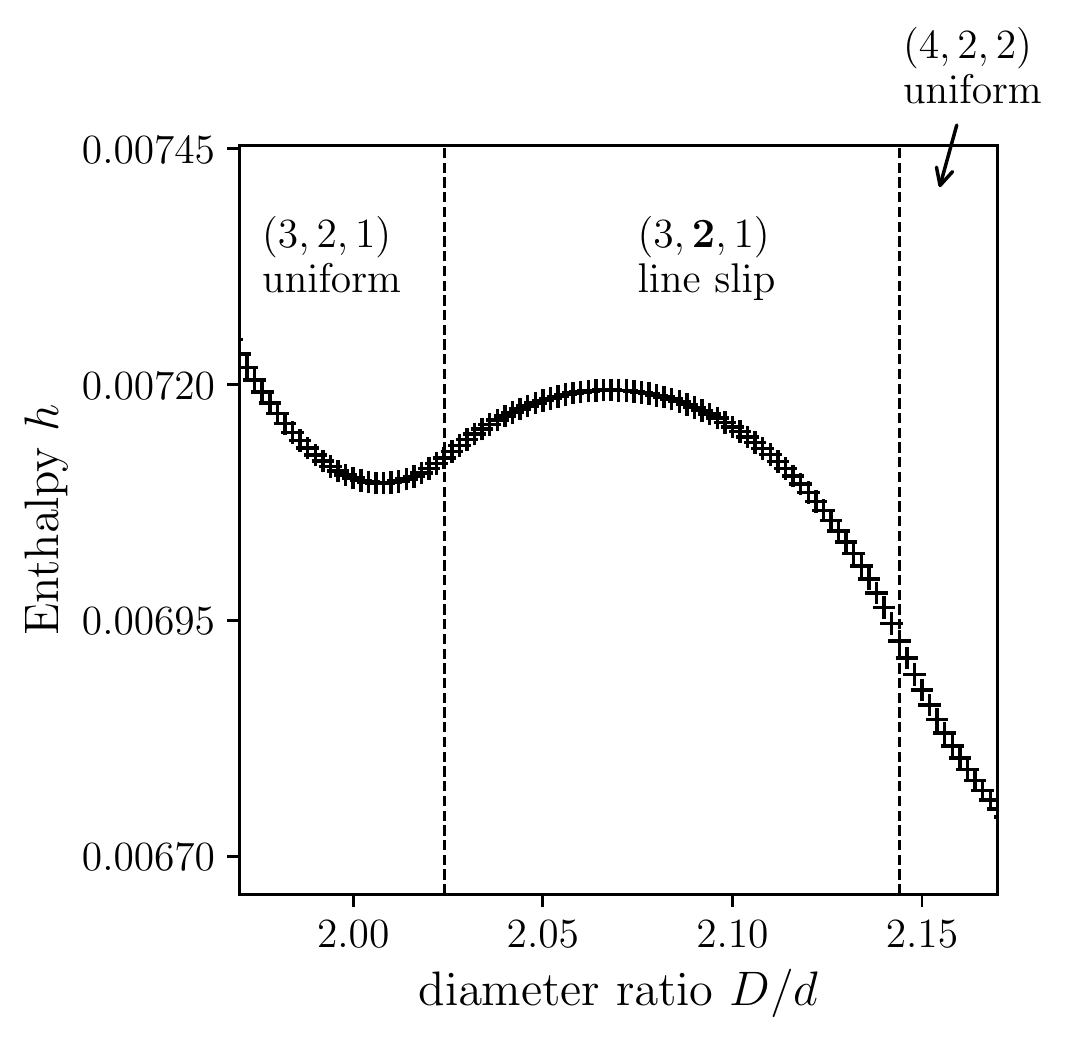}
\caption{
Variation in enthalpy $h$ as a function of $D/d$ for pressure
$p=0.007$. The change in $D/d$ allows the packing to
continuously (and reversibly) transform between the $(3,2,1)$ uniform
structure, the  $(3,{\bf 2},1)$ line slip and the $(4,2,2)$ uniform
structure. The values of $D/d$ at which the nature of the structure changes are indicated by the dashed vertical lines.
}
\label{enthalpy321a}
\end{center}
\end{figure}

An example of a computed enthalpy curve is shown in Fig.
\ref{enthalpy321a}. It shows the change in enthalpy as $D/d$ is increased
and pressure is held constant at $p=0.007$. At such a low pressure a change in $D/d$
allows the $(3,2,1)$ uniform structure to transition continuously into the
$(3,{\bf 2},1)$ line-slip arrangement (by the loss of a contact),  and then
continuously into the $(4,2,2)$ uniform structure (by the formation of a
new contact), both reversibly. The values of $D/d$ where the changes in the nature of the structure occur are indicated by dashed vertical lines in Fig. \ref{enthalpy321a}. The smooth change in the enthalpy over the course of these transitions demonstrates that the process is continuous and reversible.
The smooth change in structure, together with its rolled-out contact
network and its position in the stability diagram, Fig
\ref{hysteresis321}(b), which we will explain in more detail in Sec.{}\ref{metastability}B, can be observed in video S1 \cite{supplemental}.


\begin{figure}[t]
\begin{center}
\includegraphics[width=0.99\columnwidth]{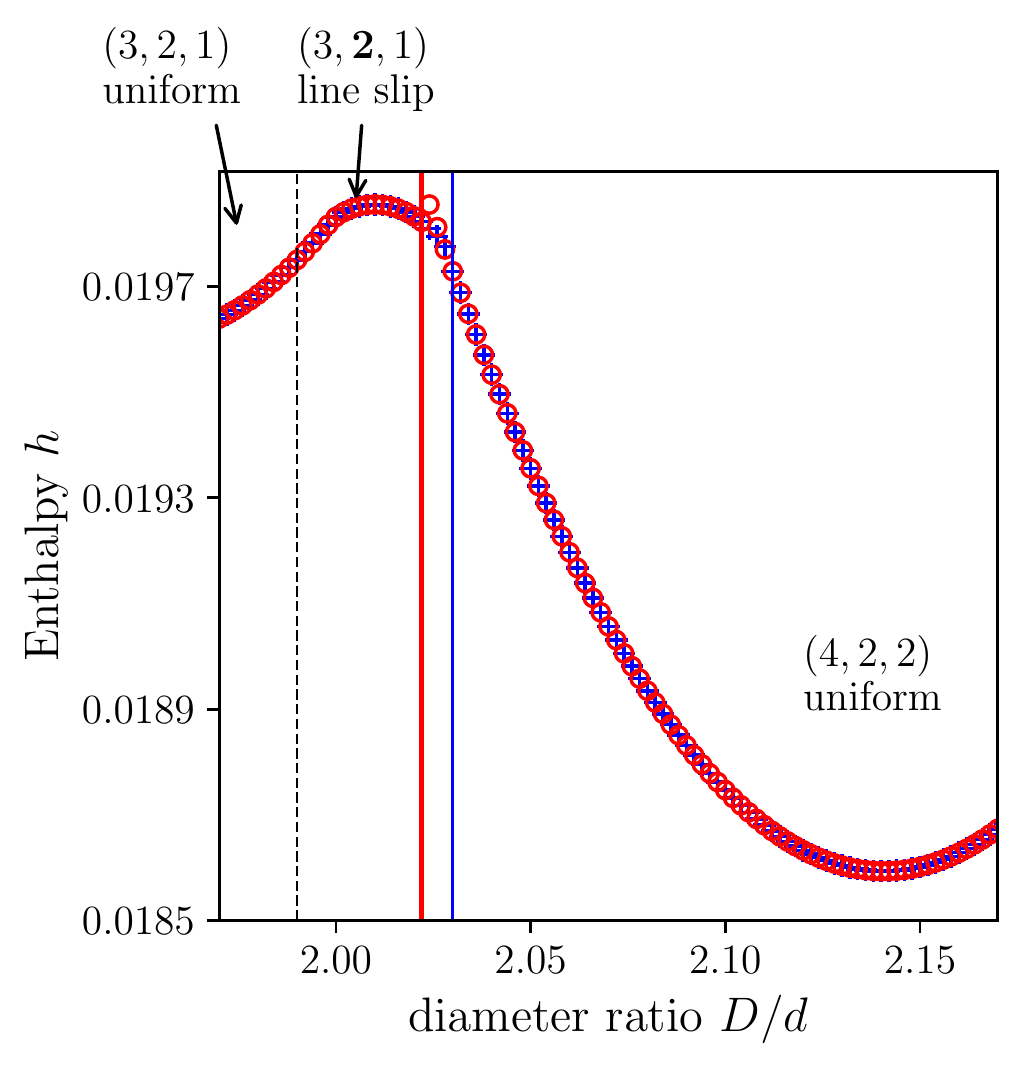}
\caption{
Variation in enthalpy $h$ as a function of $D/d$ for pressure $p=0.020$.
Changing $D/d$ leads to a reversible and continuous transformation between
the $(3,2,1)$ uniform structure and the $(3,{\bf 2},1)$ line slip for both
the forward (increasing $D/d$,  \textcolor{blue}{blue} crosses) and reverse (\textcolor{red}{red} circles)
trajectories, indicated by the vertical dashed line. In contrast, the transition from the $(3,{\bf 2},1)$
line-slip structure to $(4,2,2)$ uniform arrangement (thin \textcolor{blue}{blue} line) is discontinuous and
occurs at a lower value of $D/d$ on the reverse trajectory (thick \textcolor{red}{red} line). 
}
\label{enthalpy321b}
\end{center}
\end{figure}

At the higher pressure of $p = 0.02$ the nature of the transitions between
these structures is different, as illustrated in Fig. \ref{enthalpy321b}.
As before, and shown by the blue crosses, starting with the $(3,2,1)$
uniform structure, a steady increase in $D/d$ leads to a smooth change in
the enthalpy leading to the $(3,{\bf 2},1)$ line slip at a value of $D/d$,
as indicated by the vertical blue dashed line.
However increasing $D/d$ further leads eventually to a discontinuous transition whereby the structure transforms suddenly into the $(4,2,2)$ uniform arrangement, as indicated by the continuous vertical blue line. 
At this point in $D/d$ the enthalpy shows a sudden drop.
Video S2 \cite{supplemental} shows the varying structures along this trajectory and the sudden transformation into the (4, 2, 2) uniform arrangement.

When the diameter ratio is decreased again (red circles), the transition from the $(4,2,2)$ uniform structure to the $(3,{\bf 2},1)$ line slip is again discontinuous (thick red vertical line). However, the transition to the $(3,{\bf 2},1)$ line-slip structure occurs at a \emph{lower} value of $D/d$ on the reverse trajectory as compared to the forward trajectory and thus exemplifies the structural hysteresis present in these packings above a critical pressure.

\begin{figure}[t!]
\begin{center}
\includegraphics[width=0.99\columnwidth]{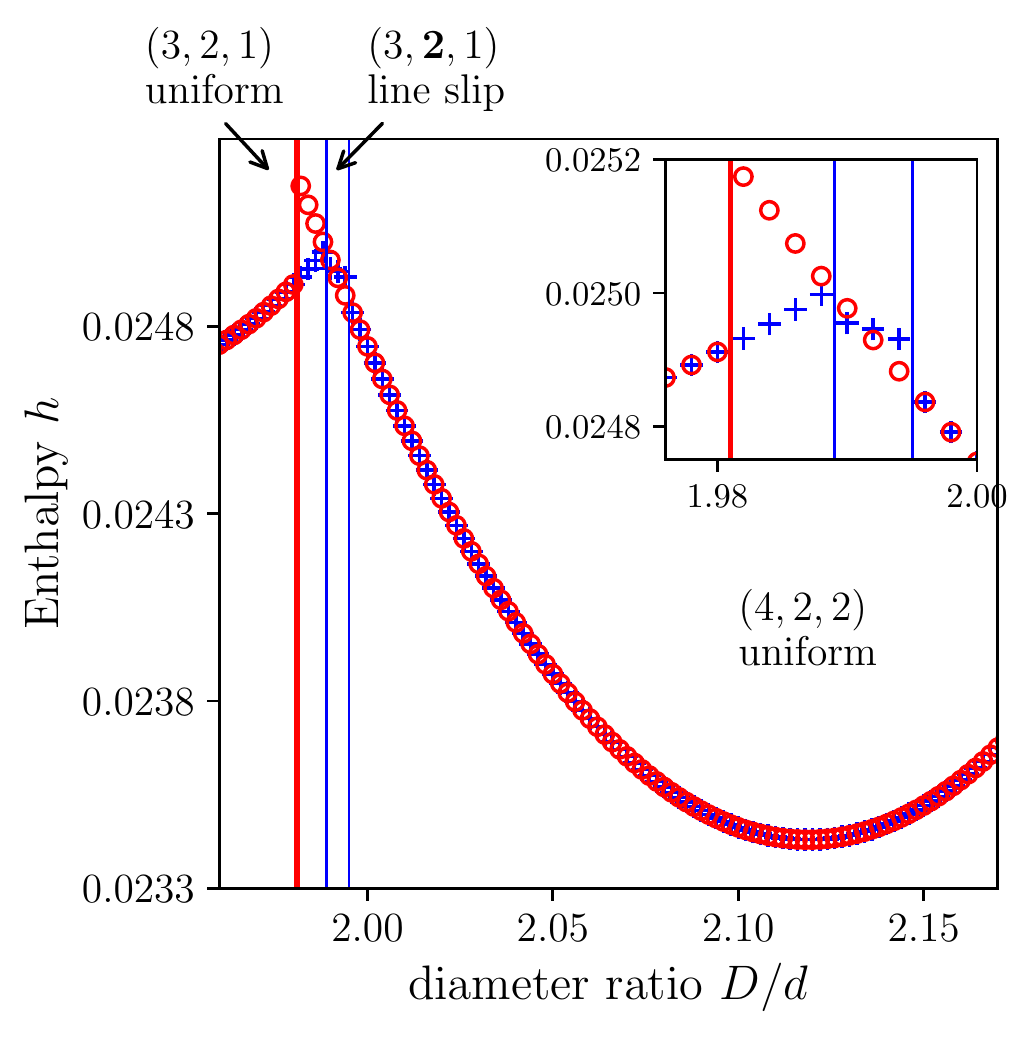}
\caption{
Variation in enthalpy $h$ as a function of $D/d$ while holding pressure at
a constant value of $p=0.026$. The forward trajectory is as before: an
increase in $D/d$ leads to a discontinuous transformation from the $(3,2,1)$
uniform structure to the $(3,{\bf 2},1)$ line-slip (thin \textcolor{blue}{blue} line), a
further increase in $D/d$ results in a discontinuous transition to the
$(4,2,2)$ uniform structure (thin \textcolor{blue}{blue} line). The reverse trajectory is
remarkable in that the intervening line-slip is eliminated. Instead the transition from the $(4,2,2)$ uniform structure to the $(3,2,1)$ uniform structure is via a discontinuous transition (thick \textcolor{red}{red} line).
The inset shows a zoom on the discontinuous transitions.
}
\label{enthalpy321c}
\end{center}
\end{figure}

At even greater pressures the line-slip structure disappears completely from the reverse trajectory, as shown in Fig. \ref{enthalpy321c} for $p = 0.026$.
Now, in the forward trajectory  a discontinuous transition (vertical thin blue line) transforms the
$(3,2,1)$ uniform packing into the $(3,{\bf 2},1)$ line slip and then to
the $(4,2,2)$ uniform structure by a further discontinuous transition.
On the reverse trajectory the $(4,2,2)$ uniform structure jumps straight to the $(3,2,1)$ uniform structure by a discontinuous transition (vertical thick red line) - without the presence of the intervening line slip.
Increasing the pressure yet higher ($p \lesssim 0.028$) eliminates the line slip also on the forward trajectory so that transformations between the $(3,2,1)$ uniform structure and the $(4,2,2)$ uniform structure are accompanied only by discontinuous transitions, and the line slips are completely eliminated.
We display again the structures and rolled-out contact networks along these trajectories in videos S3 and S4 \cite{supplemental}.

\subsection{Stability diagrams}
\label{ss:stability}

\begin{figure*}
\begin{center}

\begin{subfigure}[b]{0.48\textwidth}
\includegraphics[width=1.0\textwidth]{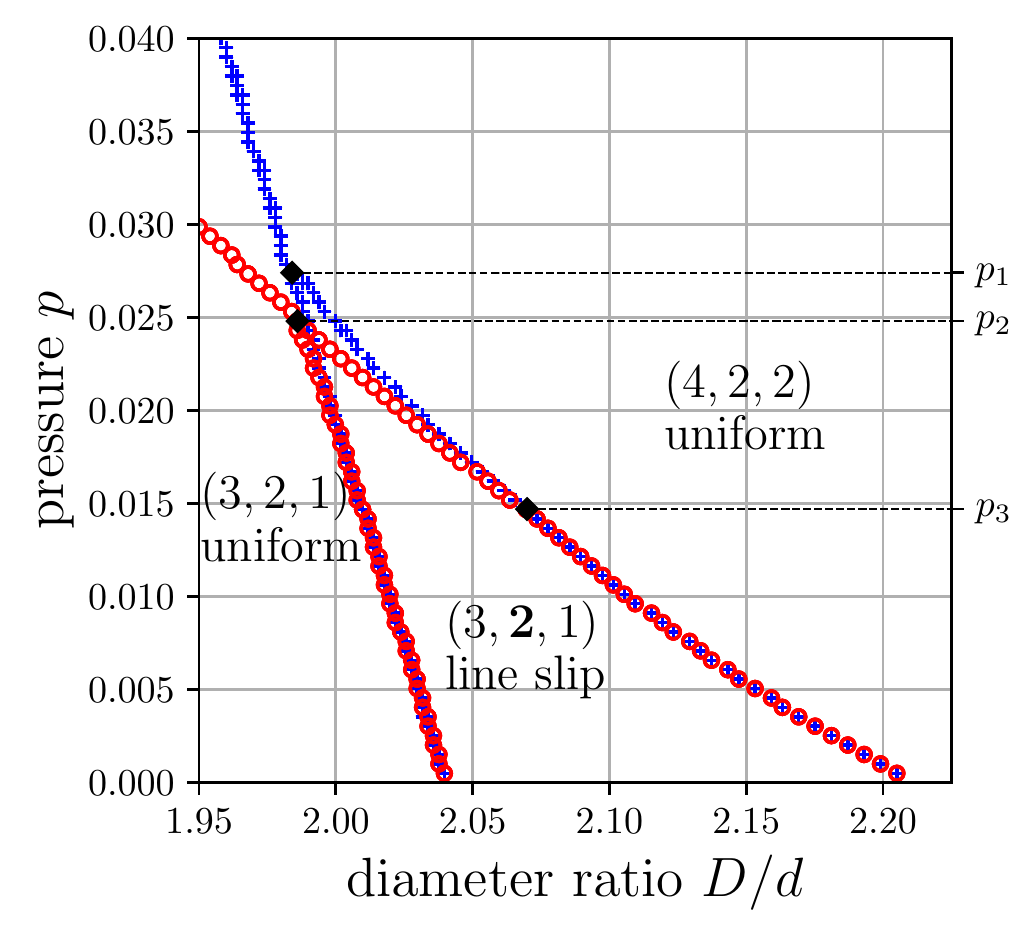}
\caption{}
\end{subfigure}
\begin{subfigure}[b]{0.48\textwidth}
\includegraphics[width=0.95\textwidth]{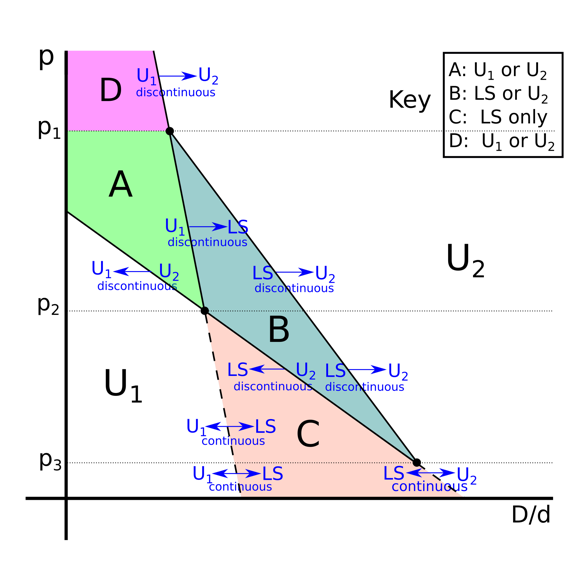}
\caption{}
\end{subfigure}
\caption{
(a) The stability diagram for transitions between $(3,2,1)$, $(4,2,2)$ and the associated line slip $(3,{\bf 2},1)$. The computed transition points are indicated by blue crosses (increasing $D/d$) and red circles (decreasing $D/d$) according to the direction taken.
This is indicated by arrows in the accompanying schematic diagram, (b).
Here the two uniform arrangements are labeled $U_1$ and $U_2$ and the
intermediate line-slip arrangement is labeled LS.  This diagram
represents the topology of (a), but does not preserve geometrical features. Continuous transitions are marked by dashed lines, and discontinuous transitions are represented by solid lines.
}
\label{hysteresis321}
\end{center}
\end{figure*}

The stability diagram represents the boundaries at which transitions take place between a particular set of structures (starting with one of the structures listed in it).
Other metastable structures must exist in the same region, but are not represented.
As mentioned earlier, the stability diagram is not to be confused with the
phase diagram in Fig \ref{phasediagram}. 

Figure \ref{hysteresis321}(a) is the stability diagram representing transitions between the $(3,2,1)$ and $(4,2,2)$ uniform structures and the associated line slip $(3,{\bf 2},1)$.
It was obtained from the calculations of enthalpy curves of the kind
described in Sec.{}\ref{metastability}B.
Figure \ref{hysteresis321}(b) is a schematic guide to the interpretation of this stability diagram, which is correct in representing the topological features of the stability diagram but does not preserve the geometrical features.
Here $U_1$ is to be identified with the $(3,2,1)$ uniform packing, $U_2$ is to be identified with the $(4,2,2)$ uniform packing and LS with the $(3,{\bf 2},1)$ line slip.

Continuous transitions (reversible) between structures are marked by dashed
lines while discontinuous transitions (irreversible) are represented by
solid lines.
Blue arrows indicate the directions for which such boundaries entail a
transition.  
We also mark the nature of the boundary and the structures encountered on either side of the boundary. 

The reversible boundaries are to be identified with parts of the phase boundaries of the equilibrium phase diagram, Fig. \ref{phasediagram}. 

As displayed in Fig. \ref{hysteresis321}, there are four qualitatively
different pressure regimes, corresponding to the examples described above.
These are $p < p_3$ (see Fig. \ref{enthalpy321a}), $p_3 < p < p_2$ (see Fig. \ref{enthalpy321b}), $p_2 < p < p_1$ (see Fig. \ref{enthalpy321c}) and $p_1 < p$.
For the last regime, i.e. above $p_1$, the metastable phase of the $(3, \bm{2}, 1)$ line slip has completely vanished.
Here the enthalpy curves show discontinuities at the blue crosses for increasing $D/d$ and at the red crosses for decreasing $D/d$.
Videos S1-S4 \cite{supplemental} illustrate the change in structure for all four different pressure regimes when varying $D / d$ at constant pressure.
At the discontinuous transitions the corresponding videos reveal the sudden change in structure.

To illustrate the useful application of the schematic diagram Fig. \ref{hysteresis321}(b) we describe the case $p_3 < p < p_2$ in detail by choosing $p=0.020$, which corresponds to the enthalpy curves shown in Fig. \ref{enthalpy321b}. The other cases can be interpreted similarly by comparing Fig.{}\ref{hysteresis321}(a) with Fig.{}\ref{hysteresis321}(b).

Starting from the $(3,2,1)$ uniform packing at $p=0.020$ increasing $D/d$ leads to a boundary [shown by the blue crosses in Fig. \ref{hysteresis321}(a)] which marks the continuous transition to the $(3,{\bf 2}, 1)$ line slip. In Fig. \ref{hysteresis321}(b) this is marked by the dashed line indicating the continuous transition $U_1 \leftrightarrow LS$. 

Increasing $D/d$ further a second boundary (blue crosses) is encountered, making a discontinuous transition to the $(4,2,2)$ uniform structure. This boundary is shown in the schematic diagram by the solid line and is labeled $LS \rightarrow U_2$. The arrow indicates that this is encountered only upon increasing $D / d$, transforming the line slip $LS$ into the uniform structure $U_2$ by a discontinuous transition. 

Upon decreasing $D/d$ the $(4,2,2)$ uniform structure undergoes a discontinuous transition at the boundary indicated by red dots in Fig \ref{hysteresis321}(a) into the $(3,{\bf 2},1)$ line slip.
This boundary is marked by the solid line in the schematic diagram accompanied by the label $LS\leftarrow U_2$. 

A further decrease of $D / d$ results in a continuous transition into the uniform $(3, 2, 1)$ structure, at the boundary marked by red dots in Fig \ref{hysteresis321}(a) and dashed line in Fig \ref{hysteresis321}(b).
We have thus returned to the starting point of this particular excursion
through the stability  diagram.

\section{Conclusion}
We have not attempted an exhaustive description of metastability and
hysteresis in this system, as this seems impractical. Simulations in other
regions of the phase diagram have resulted in further stability diagrams,
most of which are qualitatively similar to Fig. \ref{hysteresis321}, as
might be expected from the repetitive pattern of the phase diagram , Fig.
\ref{phasediagram}. Thus, Fig. \ref{hysteresis321}(b), given the coordinates
of about six points, where the lines
meet can serve as a guide to any such case. A few others are different; all of  these will be published in due course.


It is perhaps surprising that all of the simulations resulted only in the
appearance of structures that are to be found in the equilibrium phase
diagram, albeit over different ranges of $p$ and $D/d$. This may be
rationalised at or close to the hard sphere limit, but might not have been
anticipated at higher pressures. It would appear that the procedure adopted
here allowed the appearance of radically new structures, but none were found.
More extended simulations may yet turn up surprises.

It may be interesting to extend this analysis to higher values of $D/d$,
for which Fu {\em et al.} \cite{fu2016hard} have computed a list of
equilibrium structures that are of a different character, as we noted at
the outset. This is likely to be quite demanding, and should perhaps be
guided by preliminary experimental investigations of that regime.

The results should provide insight for new experiments in which it is more convenient to vary $p$ for fixed $D/d$, rather than the reverse. For example, ordered columnar bubble structures are readily formed by simple procedures \cite{winkelmann2017simulation}, and there is a natural variation of $p$ within the column in this experiment, due to gravity. Such a system may be used to identify structures and transitions between them, for comparison with what is presented here.   
Diagrams such as Fig \ref{hysteresis321}(b) may then serve as a guide to the practical
fabrication of structures of soft spheres in tubes, for which the equilibrium
phase diagram by itself, may be misleading.
\newpage
 
\acknowledgments
D.W. and A.M. would like to thank the Isaac Newton Institute for Mathematical Sciences for support and hospitality during the programme Growth Form and Self-Organisation  when work on this paper was undertaken. This work was supported by: EPSRC Grant No. EP/K032208/1 and EP/R014604/1, as well as an Irish Research Council Postgraduate Scholarship (project ID GOIPG/2015/1998). We also acknowledge the support of the Science Foundation Ireland (SFI) under Grant No. 13/IA/1926 and MPNS COST Action MP1305 `Flowing Matter' and the European Space Agency ESA MAP Metalfoam (AO-99-075) and Soft Matter Dynamics (contract: 4000115113)
A.M.'s visit to the TCD Foams and Complex System Group was funded by the Visiting Professorships and Fellowships Benefaction Fund.

The authors would also like to acknowledge numerous useful discussions with B. Haffner.

\bibliographystyle{nonspacebib}

\end{document}